\documentclass{aa}
\usepackage{graphicx}
\usepackage{txfonts}
\usepackage{natbib}
\begin{document}
   \title{X-ray sources and their optical counterparts in the globular cluster \object{M 22}}

   \subtitle{}

   \author{N.A. Webb
          \inst{1}
          \and
          D. Serre
          \inst{1}
          \and
          B. Gendre
          \inst{1,4}\thanks{Present address}
          \and 
          D. Barret
          \inst{1}
          \and 
          J.-P. Lasota
          \inst{2}
          \and 
          L. Rizzi 
          \inst{3}
          }

   \offprints{N.A. Webb \email{Natalie.Webb@cesr.fr}}

   \institute{Centre d'Etude Spatiale des Rayonnements, 9 avenue du Colonel Roche, 31028 Toulouse Cedex 04, France 
    \and           
        Institut d'Astrophysique de Paris, 98bis boulevard Arago, 75014 Paris, France 
    \and           
        INAF, Osservatorio Astronomico di Padova, Vicolo dell'Osservatorio 5, I-35122 Padua, Italy 
    \and
       Istituto di Astrofisica Spaziale e Fisica Cosmica, C.N.R., Via Fosso del Cavaliere, Roma, Italy
         }

   \date{Received 8 March 2004; accepted 26 May 2004}

   \abstract{Using XMM-Newton EPIC imaging data, we have detected 50
   low-luminosity X-ray sources in the field of view of M~22, where
   5$\pm$3 of these sources are likely to be related to the cluster.
   Using differential optical photometry, we have identified probable
   counterparts to those sources belonging to the cluster.  Using
   X-ray spectroscopic and timing studies, supported by the optical
   colours, we propose that the most central X-ray sources in the
   cluster are cataclysmic variables, millisecond pulsars, active
   binaries and a blue straggler.  We also identify a cluster of
   galaxies behind this globular cluster.   }

\authorrunning{Webb et al.}
\titlerunning{X-ray/optical sources in M 22}

   \maketitle
%

\section{Introduction}

It is expected that globular clusters (GCs) should contain many binary
systems, due to interactions occurring within the clusters 
\citep[e.g.][]{dist94,rasi00,port97}, and that these systems could
play a critical role in the dynamical evolution of GCs, serving as an
internal energy source which counters the tendency of cluster cores to
collapse \citep[see][for a review]{hut92}.  However, these binaries
are difficult to locate, because of crowding in optical observations.
The binaries, which are also visible at high energies, can be located
using X-ray observations, where the crowding is less severe.  Indeed
the small population of bright X-ray sources in globular clusters
(L$_x >$ 10$^{36}$ erg s$^{-1}$), known to be X-ray binaries
\citep{hert83}, were detected primarily through their X-ray
bursts.  However, there is also a population of low-luminosity
\citep[L$_x\ _\sim ^<$ 10$^{34.5}$ erg s$^{-1}$][]{hert83,verb01} X-ray
sources.  Thanks to the new generation of X-ray observatories,
e.g. XMM-Newton and Chandra, the number of sources belonging to this
population has rapidly increased.  The nature of most these faint
X-ray sources can only be determined using follow-up optical
observations.  A variety of objects have been identified, including
many binary systems, such as: X-ray binaries
\citep[e.g.][]{gend03b,rutl02}; cataclysmic variables
\citep[e.g.][]{cars00,gend03a}; millisecond pulsars
\citep[e.g.][]{grin01,cami00}; active binaries
\citep[e.g.][]{kalu96,cars00}; as well as some fore- and background
objects, e.g. stars \citep[e.g.][]{gend03a} or clusters of galaxies
\citep[e.g.][]{hert83}.

As a result of these recent identifications, we are now also able to
start to constrain the evolutionary paths of these binaries in GCs
\citep[e.g. the formation of neutron star low mass X-ray binaries][]{gend03b,pool03} and thus not only begin to comprehend the evolution of these 
systems, but also understand the role played by binaries in countering
the core-collapse of GCs.  We have therefore obtained X-ray data of
the central 30\arcmin\ of the globular cluster \object{M 22}
(\object{NGC 6656}) with {\it XMM-Newton} to detect the low-luminosity
X-ray sources in this globular cluster.  This GC is one of the closest
\citep[2.6 $\pm$ 0.3 kpc][]{pete94} and thus it has
previously been chosen as a target by the X-ray satellites {\it Rosat}
and {\it Einstein} (see e.g. Verbunt 2001 and references therein).
Eight X-ray sources \citep[L$_x\ _\sim ^<$ 10$^{34.5}$ erg
s$^{-1}$][]{hert83}, were detected in the direction of the cluster
\citep{john94}, using the {\it Rosat} PSPC.  The source
detected within the core radius (X4/B) was associated with the cluster.  This
source has been noted to be somewhat variable, varying by a factor 3
in flux between {\it Rosat} observations.  The eight sources detected
by {\it Rosat} include two of the four X-ray sources detected by {\it
Einstein} \citep{hert83}.  Using the {\it XMM-Newton} MOS data only,
we detected a total of 34 sources
\citep{webb02}, including three sources within the core radius, where
at least two of these are likely to be associated with the cluster.  
Here we present both the MOS and the PN data.  

There have been several optical observing campaigns carried out on
this cluster, however there has been very little photometric data
taken at the blue end of the spectrum. \cite{piet03,kalu01,clem01}
have identified a variety of variable stars in \object{M~22},
including SX Phe stars, RR Lyr stars and candidate eclipsing binaries.
\cite{mona04} have recently carried out wide field
photometry in the B-, V- and I-bands of \object{M~22}, which they have
presented in conjunction with some H$_\alpha$ data and J-, H- and K-band data
from the 2 MASS survey to characterise the evolved stellar sequences.
They also identify on their colour-magnitude diagrams, previously
identified variable stars in this globular cluster. \cite{ande03} have
recently identified a new variable star in the centre of
\object{M~22}.  From its variability, its H$_\alpha$ emission and its
coincidence with the {\it Einstein}/{\it Rosat} source X4/B, they
conclude that this source is one of a very small number of confirmed
or probable dwarf nova eruptions seen in globular clusters and the
first to be found in such a low-concentration cluster.  

Using the good astrometry of the X-ray sources, we use differential
optical photometry (U, B and V filters) of the same field of view as
the X-ray data, to identify blue (hot) sources that fall within the
error circle of the the X-ray sources.  Optical sources that are bluer
than the main sequence stars of the cluster are likely to be
interacting binaries but also cooling white dwarfs, millisecond
pulsars or blue stragglers and thus the optical counterparts of the
X-ray sources.  Also, taking advantage of the spectral and timing
information of the brightest X-ray sources we determine the nature of
a variety of low-luminosity X-ray sources in the field of
\object{M 22}.

\section{Observations and data reduction}

\subsection{X-ray data}

We obtained 37 kiloseconds (ks) of EPIC MOS and 34 ks of EPIC PN data
of the globular cluster \object{M 22} (\object{NGC 6656}), with {\it
XMM-Newton}.  However, 14 ks were affected by high background activity
(soft proton flare).  Observations were made on September 19-20 2000 ,
during the `Routine Observing Phase', in the full frame mode
\citep{turn01}, with the medium filter.  Some of these data (MOS) 
have been presented previously in \cite{webb02}, but due to a problem
with the `Attitude History File', the PN data could not be processed.
This problem has now been rectified and we present both the MOS and
the PN data, reduced with Version 5.4.1 of the {\it XMM-Newton} SAS
(Science Analysis Software).

The MOS data were reduced using `emchain' with `embadpixfind' to
detect the bad pixels.  The event lists were filtered, so that 0-12 of
the predefined patterns (single, double, triple, and quadruple pixel
events) were retained and the high background periods were identified
by defining a count rate threshold above the low background rate and
the periods of higher background counts were then flagged in the event
list.  We also filtered in energy. We used the energy range 0.2-10.0
keV, as recommended in the document `EPIC Status of Calibration and
Data Analysis' \citep{kirs02}.  

The PN data were reduced using the `epchain' of the SAS.  Again the
event lists were filtered, so that 0-4 of the predefined patterns
(single and double events) were retained, as these have the best
energy calibration.  We also filtered in energy, where we used the
energy range 0.5-10.0 keV.

\begin{figure}
   \centering \includegraphics[angle=0,width=8.5cm]{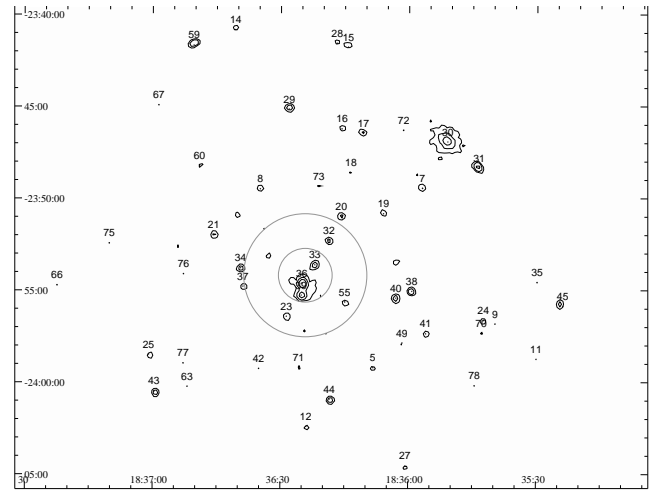}
\caption{X-ray contour plot of the central region of M~22. The inner circle shows the core radius
   and the outer circle shows the half mass radius.  The abscissa
   indicates the right ascension in $^h$ \hspace*{1mm} $^m$
   \hspace*{1mm} $^s$ and the ordinate indicates the declination in
   $^{\circ}$ \hspace*{1mm} \arcmin \hspace*{1mm} \arcsec.  The
   contours represent 5, 7 and 14 $\sigma$ confidence levels and the
   source numbers are the same as those found in
   Table~\ref{tab:xsources}. }
\label{fig:m22contours}
\end{figure}

The source detection was done in the same way as
\cite{gend03a,gend03b}, but briefly we used the SAS wavelet detection
algorithm on the 0.5-5.0 keV image, as the signal-to-noise ratio was
best in this range.  We kept only those sources detected with two or
more cameras, with the exception of sources found with the PN camera,
which is more sensitive than the two MOS cameras.  To improve the
astrometry, we used the 3 sources that were detected both in our data
and by the {\it Rosat} HRI (sources 36, 30 and 60,
Table~\ref{tab:xsources}).  An IDL routine was used to derive the
adjustment required for the MOS data to align with the HRI data.  We
found that we required both a small transversal shift and rotation, in
the same manner as Hasinger et al (2001).  This leads to a residual
error of $\sim$5.6'', where the largest error is due to the position
error of the sources detected by the HRI.  We detected 50 sources with
a maximum likelihood greater than 4.5$\sigma$, which are given in
Table~\ref{tab:xsources}, along with their position and count rate.
We also give the identification number for the sources detected with
only the MOS cameras in \cite{webb02} and the sources detected by {\it
Einstein} and {\it Rosat}.  The positional error is the 90\%
confidence mean statistical error on the position of the {\it
XMM-Newton} sources convolved with the mean statistical error on the
position of the {\it Rosat} sources \citep{verb01}.  The majority of
these sources can be seen in the contour plot of the central regions
of \object{M~22}, see Fig.~\ref{fig:m22contours}.

\subsection{Optical data}

Schott glass U-, B- and V-filter photometry was taken with the Wide
Field Imager (WFI) mosaic at the 3.9m {\it Anglo Australian Telescope}
(AAT) on September 15-16 2001.  The camera consists of eight 2000
$\times$ 1000 pixel CCDs, in a square array of 2 $\times$ 4 detectors,
which are mounted at the triplet-corrected f/3.3 prime focus of the
AAT on the {\it prime focus unit}, which results in a field of view
(FOV) of 33\arcmin $\times$ 33\arcmin, similar to the {\it XMM-Newton}
FOV. Sky conditions were photometric on the second night.  Standard
star fields taken from \cite{land92} were also observed on 2001
September 13 for calibration purposes.

Each image was bias-subtracted and then flat-fielded with twilight sky
flats, using the {\it IRAF} software \citep{tody86,tody93}. The long
U-band observations were also dark subtracted, as the dark-current was
non-negligible in these longer exposures.  The data were
astrometrically calibrated using the {\it IRAF} package {\it MSCRED}
\citep{vald98} and the package {\it Wide Field Padova REDuction}
(WFPRED) developed at the Padova Astronomical Observatory (Held et
al. in preparation) and the standard star fields.  A refinement to the
astrometry was then made using the UCAC (USNO CCD Astrograph Catalog)
\citep{zach00} of the
\object{M~22} data, so that the positions are good to less than
0.1\arcsec.  The data for each filter were then stacked to create a
deep image and the photometry was carried out using DAOPHOT/ALLSTAR
\citep{stet87}. The instrumental magnitudes were corrected using the 
standard stars.  We find 123\hspace*{0.5mm}220 stars detected in the
U-band and at least one of the other two bands, where the maximum
matching distance was 3 pixels ($<1.5$\arcsec).  The magnitudes range
from approximately 14.0 to 22.8 in each filter. The photometric errors
range from 0.01 to 0.1 magnitudes.

\begin{table}
  \caption[]{X-ray sources in the direction of M22, as determined from the EPIC observations, 0.5-10.0 keV band.  The identification number, any former identifications and position are given, along with the error on this position and the count rate.}
     \label{tab:xsources}
       \begin{tabular}{lccccc}
         \hline
         \noalign{\smallskip}
           & For. & RA (2000) & Dec (2000) & Error & Count s$^{-1}$ \\
           ID & ID & $^h$ \hspace*{3mm} $^m$ \hspace*{3mm} $^s$ & $^{\circ}$ \hspace*{3mm} ' \hspace*{3mm} '' & '' & $\times10^{-3}$ \\
         \hline
 5& 24 & 18\ 36\  08.28&  -23\ 59\  13.12& 6.04&  6.09$\pm$ 1.04\\
 7& 6 & 18\ 35\  56.82&  -23\ 49\  27.69& 6.63&  5.30$\pm$0.89\\
 8& 5 & 18\ 36\  34.79&  -23\ 49\  29.62& 6.24&  11.65$\pm$ 1.29\\
 9& & 18\ 35\  39.66&  -23\ 56\  50.22& 7.99&  4.73$\pm$0.98\\
11& & 18\ 35\  29.99&  -23\ 58\  44.97& 7.02&  6.40$\pm$ 1.14\\
12& 1 &18\ 36\  23.97&  -24\ 02\  24.67& 6.16&  8.30$\pm$ 1.21\\
14&  &18\ 36\  40.67&  -23\ 40\  50.49& 6.22&  20.66$\pm$ 2.56\\
15& b &18\ 36\  14.28&  -23\ 41\  47.36& 5.78&  22.94$\pm$ 2.94\\
16& 3 &18\ 36\  15.69&  -23\ 46\  13.76& 6.44&  4.59$\pm$0.75\\
17& 4,1$^R$ &18\ 36\  10.69&  -23\ 46\  29.10& 5.58&  13.14$\pm$ 1.33\\
18&  &18\ 36\  13.58&  -23\ 48\  37.03& 6.65&  3.19$\pm$0.64\\
19& 8  &18\ 36\  05.91&  -23\ 50\  50.69& 5.92&  7.74$\pm$ 1.02\\
20& 9 &18\ 36\  15.80&  -23\ 50\  58.84& 6.09&  4.77$\pm$0.61\\
21& 10 &18\ 36\  45.70&  -23\ 51\  59.16& 5.62&  10.44$\pm$ 1.04\\
23$^h$& 21  &18\ 36\  28.61&  -23\ 56\  25.92& 5.89&  6.25$\pm$0.83\\
24&  &18\ 35\  42.39&  -23\ 56\  38.67& 5.94&  12.66$\pm$ 1.80\\
25& 23 &18\ 37\  00.91&  -23\ 58\  29.39& 5.79&  11.90$\pm$ 1.49\\
27& f,7$^R$ &18\ 36\  00.62&  -24\ 04\  33.16& 6.00&  13.95$\pm$ 2.32\\
28&  &18\ 36\  16.91&  -23\ 41\  34.96& 6.22&  19.07$\pm$ 2.42\\
29& 2 &18\ 36\  28.08&  -23\ 45\  09.49& 5.42&  25.53$\pm$ 1.93\\
30& c,3$^R$,A &18\ 35\  50.91&  -23\ 46\  56.29& 5.06&  79.11$\pm$ 3.71\\
31& d &18\ 35\  43.55&  -23\ 48\  21.60& 5.13&  46.10$\pm$ 2.71\\
32$^h$& 11  &18\ 36\  18.67&  -23\ 52\  19.06& 5.51&  8.08$\pm$0.88\\
33$^c$& 13 &18\ 36\  21.97&  -23\ 53\  37.65& 5.37&  12.03$\pm$ 1.29\\
34& 14 &18\ 36\  39.37&  -23\ 53\  47.47& 5.47&  9.23$\pm$0.92\\
35&  &18\ 35\  29.84&  -23\ 54\  34.33& 8.79&  2.73$\pm$ 1.03\\
36$^c$& 16,4$^R$,B &18\ 36\  24.97&  -23\ 54\  38.04& 4.98&  49.14$\pm$ 1.81\\
37& 15 &18\ 36\  38.80&  -23\ 54\  47.68& 6.26&  3.51$\pm$0.52\\
38& 17  &18\ 35\  59.23&  -23\ 55\  04.22& 5.31&  16.19$\pm$ 1.31\\
39$^c$& 18 &18\ 36\  25.12&  -23\ 55\  16.97& 5.06&  27.73$\pm$ 1.36\\
40& 20 &18\ 36\  02.96&  -23\ 55\  26.42& 5.42&  17.92$\pm$ 1.56\\
41& 22 &18\ 35\  55.93&  -23\ 57\  22.44& 6.09&  10.31$\pm$ 1.84\\
42&  &18\ 36\  35.08&  -23\ 59\  16.15& 7.83&  0.78$\pm$0.31\\
43& 25  &18\ 36\  59.61&  -24\ 00\  29.76& 5.37&  21.46$\pm$ 1.77\\
44& 19 &18\ 36\  18.46&  -24\ 00\  56.14& 5.32&  16.03$\pm$ 1.35\\
45& g,8$^R$ & 18\  36\  27.71 & -24\ 6\ 39.72 & 7.11 & 6.33$\pm$1.14\\
49&  &18\ 36\  01.47&  -23\ 57\  52.70& 7.07&  5.26$\pm$ 1.03\\
55$^h$&  & 18\ 36\ 14.94& -23\ 55\ 38.91& 7.14 & 1.92$\pm$0.38\\
59& a  &18\ 36\  50.40&  -23\ 41\  40.54& 5.28&  64.36$\pm$ 3.80\\
60&  27,9$^R$ &18\ 36\  49.05&  -23\ 48\  13.43& 6.02&  8.27$\pm$ 1.24\\
63&  &18\ 36\  51.85&  -24\ 00\  14.54& 7.86&  1.52$\pm$0.52\\
66&  &18\ 37\  22.41&  -23\ 54\  43.35& 9.54&  6.14$\pm$ 1.61\\
67&  &18\ 36\  58.62&  -23\ 44\  55.76& 8.41&  6.78$\pm$ 1.73\\
70&  &18\ 35\  42.87&  -23\ 57\  20.08& 8.19&  5.00$\pm$ 1.37\\
71&  &18\ 36\  25.69&  -23\ 59\  11.89& 10.40& 2.58$\pm$0.73\\
72&  &18\ 36\  01.29&  -23\ 46\  18.50& 10.10& 3.66$\pm$ 1.23\\
73&  &18\ 36\  21.08&  -23\ 49\  22.20& 7.55&  2.32$\pm$0.67\\
75&  &18\ 37\  10.17&  -23\ 52\  26.63& 8.72&  3.44$\pm$ 1.13\\
76&  &18\ 36\  52.76&  -23\ 54\  07.03& 7.97&  2.60$\pm$0.72\\
77&  &18\ 36\  52.80&  -23\ 58\  58.18& 8.78&  2.62$\pm$0.83\\
78&  &18\ 35\  44.47&  -24\ 00\  11.75& 8.43&  2.56$\pm$ 1.03\\
\hline
  \end{tabular}
    \begin{list}{}{}
\vspace*{-0.3cm}
      \item $^c$ sources within the core, $^h$ sources within the half mass radius
      \item Former ID: A-B, {\it Einstein}, \cite{hert83}
      \item \hspace*{1.0cm}1-8$^R$, {\it Rosat} PSPC, \cite{john94}
      \item \hspace*{1.0cm}9$^R$, {\it Rosat} HRI, \cite{verb01}
      \item \hspace*{1.0cm}1-27, a-g, {\it XMM-Newton} MOS only, \cite{webb02}
    \end{list}
 \end{table}

\section{The X-ray sources}
\label{sec:xsources}

We have detected three sources within the core radius
\citep[85.11\arcsec][]{djor93}, sources 39, 33 and 36 (indicated by a
$^c$ in Table~\ref{tab:xsources}), and a further three sources within
the half-mass radius \citep[3.3\arcmin][]{harr99} (indicated by a $^h$
in Table~\ref{tab:xsources}).  We computed the limit of the detectable
flux within the half mass radius, assuming a 0.6 keV blackbody
spectral model in the same way as \citet{john96}. The limiting
luminosity is L$_x$ = $4 \times 10^{30}$ erg s$^{-1}$ for the PN detector
(0.5-10.0 keV) and L$_x$ = $1.2 \times 10^{31}$ erg s$^{-1}$ for the MOS
detectors at the centre of the field of view.

Many of our detected sources are background sources. We have used the
statistical Log~N-Log~S relationship of extragalactic sources derived
from the Lockman Hole \citep{hasi01} to estimate the background source
population. We converted the source count rates to fluxes using a
power law model with spectral index of -2, as \cite{hasi01}. We
employed the method of \cite{cool02} and \cite{gend03a} which takes
into account vignetting. We calculated the average limiting flux in
four annuli on the PN camera.  These annuli were bound by the core
radius, the half mass radius, 7.5\arcmin, and 15\arcmin.
We expect 1.0$\pm$0.1, 3.3$\pm$0.8, 13.9$\pm$2 and 30$\pm$3
background sources in these annuli respectively where the uncertainty
is the 10\% error estimated on the flux, as \cite{hasi01}. We find 3, 3,
17, and 27 sources in these annuli respectively.  This indicates that
5$\pm$3 sources belong to the cluster, which are grouped about the
centre.

In addition, it is also possible to calculate the probability that the
sources within the core are not simply spurious identifications with
the cluster, in the same manner as \cite{verb01}.  Using the
probability, $p$, that no serendipitous sources are detected in
the PN observation at a distance $r < R$ from the cluster centre,
located at RA=18$^h$ 36$^m$ 24.2$^s$, dec=-23$^{\circ}$ 54\arcmin\
12\arcsec\ for \object{M 22} (Djorgovski \& Meylan 1993), where $p =
1-(R/r_d)^2$ and $r_d$ is the radius of the field of view.  The
probability of not finding any sources in a single trial simply by
chance, within the core radius, is 98.9\%.  In 50 trials, for the 50
sources, the probability of finding no sources within the core radius
is 59.7\%.  It is therefore possible that the faintest source within
the core radius (source 33) may not be related to the cluster, where
from their brightness alone, the other two core sources are more
likely to be related.

\subsection{Spectral analysis}
\label{sec:xspect}

Nine X-ray sources, including all three sources within the core
radius, have enough counts that we could extract and fit a spectrum.
We extracted the spectra using an extraction radius of
$\sim$30\arcsec\ for the core sources (due to their close proximity)
and $\sim$45\arcsec\ for the rest of the sources, with the exception
of the extended source, number 30, where we used an extraction radius
of 1\arcmin.  We used a similar neighbouring surface, free from X-ray
sources to extract a background file.  We rebinned the MOS data into
15 eV bins and the PN data into 5eV bins.  We used the SAS tasks
`rmfgen' and `arfgen' to generate a `redistribution matrix file' and
an `ancillary response file', for each source.  We binned up the data
to contain at least 20 net counts/bin.  We then used Xspec (Version
11.1.0) to fit the spectra.  We initially tried simple models, such as
a power law, a blackbody, a bremsstrahlung and a raymond smith fit,
which we found to provide a good fit to the data for the majority of
the sources.  However, we found that for sources 33 and 36 and the
extended sources 29 and 30, we required a more complicated model to
fit the data.  The results of the spectral fitting can be found in
Table~\ref{tab:xspectra}, where the best fits are given, along with
the goodness of fit ($\chi^{\scriptscriptstyle 2}_{\scriptscriptstyle
\nu}$) and the flux in the 0.2-10.0 keV range.

\begin{table*}
\begin{minipage}{18cm}
\caption{In the left hand side of the table, the best fitting models to  spectra from the MOS and PN data.  Where no error value is given for the N$_H$ ($\times 10^{21}{\rm cm}^{-2}$), the N$_H$ was frozen to that of the cluster \citep[2.2$\times 10^{21}{\rm cm}^{-2}$][]{john94}.  The flux given is unabsorbed flux ($\times 10^{-14} {\rm ergs\ cm}^{-2} {\rm s}^{-1}$) in the 0.2-10.0 keV range, with errors of the order $\pm$10\%.  In the right hand side of the table, the results from testing for variability of the sources is given.  This includes the number of data bins, the length (in seconds) of each data bin and the probabilities from a Kolmogorov-Smirnov probability of constancy test and the $\chi^2$ probability of constancy test.  Source numbers correspond to those given in Table~\ref{tab:xsources} and $^m$ indicates possible cluster members, see Sect.~\ref{sec:poss_members}}
\label{tab:xspectra}
\begin{center}
\begin{tabular}{cccccccccc|ccc}
\hline
Src & N$_H$  & Model & kT (keV) &
Photon & Abundance & z & $\chi^{\scriptscriptstyle
2}_{\scriptscriptstyle \nu}$ & dof & Flux & bins & time & KS/ \\
 & $\times 10^{21}{\rm cm}^{-2}$ & & &Index  & & & & &  & & (s) & $\chi^2$ \\
\hline
\hline
36$^m$& 2.29$\pm$0.70 & PL & - & 1.70$\pm$0.13 & - & & 1.27 & 39 & 9.8 & 10 & 2000 & 9$\times10^{-3}$/\\
   & 2.2 & BB & 0.65$\pm$0.03 & - & - & & 3.11 & 40 &  & & & 3$\times10^{-5}$\\
   & 1.45$\pm$0.50 & Brems. & 9.69$\pm$3.11 & - & - & & 1.30 & 39 &  & & & \\
   & 1.50$\pm$0.50 & RS & 8.96$\pm$3.04 & - & 0.37$\pm$0.53 & & 1.36 & 37 &  & & & \\
   & 2.2 & PL+Gau.$^*$ & - & 2.03$\pm$0.21 & - & & 1.08 & 36 &  & & & \\
\hline
33$^m$ & 1.4$\pm$2.7 & PL & - & 1.98$\pm$0.62 & - & & 0.59 & 11 & 1.7 &4 & 5000& 1$\times10^{-2}$/ \\
   & 2.2 & BB & 0.48$\pm$0.08 & - & - & & 0.80 & 12 &  & & & 0.85\\
   & 1.0$\pm$2.1 & Brems. & 3.60$\pm$2.90 & - & - & & 0.58 & 11 &  & & & \\
   & 2.2 & RS + RS & 0.07+3.76$^@$ & - & - & - & 0.49 & 10 & & & & \\
\hline
39$^m$ & 2.2 & PL & - & 1.45$\pm$0.12 & - & & 1.52 & 25 & 5.6  &10 & 2000& 2$\times10^{-10}$/ \\
   & 2.2 & BB & 0.90$\pm$0.06 & - & - & & 2.80 & 25 &  & & & 3$\times10^{-4}$\\
   & 2.2 & Brems. & 19.98$\pm$13.35 & - & - & & 1.57 & 25 &  & & & \\
   & 2.2 & RS & 19.91$\pm$19.17 & - & 6.7$\pm$2.2$\times10^{-3}$ & & 1.63 & 24 &  & & & \\
\hline
32$^m$ & 2.2 & PL & - & 6.89$\pm$1.63 & - & & 2.39 & 4 & 0.6  & 4 & 5000 & 2$\times10^{-7}$/ \\
   & 2.2 & BB & 0.13$\pm$0.02 & - & - & & 2.37 & 4 &  & & & 0.85\\
\hline
43 & 5.3$\pm$4.9 & PL & - & 6.3$\pm$3.45 & - & & 0.79 & 9 & 3.1 &4 & 5000& 6$\times10^{-3}$/ \\
   & 2.2 & BB & 0.15$\pm$0.02 & - & - & & 0.86 & 10 &  & & & 0.86\\
   & 1.1$\pm$2.2 & Brems. & 0.48$\pm$0.28 & - & - & & 0.83 & 9 &  & & & \\
   & 9.3$\pm$4.0 & RS & 0.18$\pm$0.06 & - & 2$\pm$64 & 0.16$\pm$0.4 & 1.30 & 7 &  & & & \\
\hline
44 & 2.2 & PL & - & 1.56$\pm$0.20 & - & & 4.16 & 6 & 4.2  &4 & 5000& 1$\times10^{-2}$/  \\
 & & & & & & & & &  & &  & 0.85 \\
\hline
30 & 3.5$\pm$0.4 & PL & - & 2.71$\pm$0.13 & - & & 1.73 & 62 & 85.3 &10 & 2000& 2$\times10^{-2}$/ \\
   & 2.2 & BB & 0.39$\pm$0.01 & - & - & & 3.50 & 63 &  & & & 4$\times10^{-2}$\\
   & 1.9$\pm$0.3 & Brems. & 1.98$\pm$0.20 & - & - & & 1.72 & 62 &  & & & \\
   & 1.88$\pm$0.27 & RS & 2.22$\pm$0.20 & - & 0.45$\pm$017 & 0.10$\pm$0.01 & 1.38 & 60 &  & & & \\
   & 1.9$\pm$0.30 & MEKAL & 2.25$\pm$0.21 & - & 0.44$\pm$0.18 & 0.10$\pm$0.02 & 1.37 & 59  & & & \\
\hline
31 & 0.80$\pm$0.90 & PL & - & 2.48$\pm$0.44 & - & & 0.86 & 21 & 30.0  &10 & 2000& 2$\times10^{-6}$/ \\
   & 2.2 & BB & 0.23$\pm$0.02 & - & - & & 1.81 & 22 &  & & & 0.19\\
   & 2.2 & Brems. & 0.75$\pm$0.11 & - & - & & 1.25 & 22 &  & & & \\
   & 6.5$\pm$3.0 & RS & 0.77$\pm$0.16 & - & 1.00$\pm$3.6 & 1.57$\pm$0.05 & 1.70 & 19 &  & & & \\
\hline
29 & 1.70$\pm$0.70 & PL & - & 1.68$\pm$0.22 & - & & 0.74 & 18 & 16.0  &10 & 2000& 2$\times10^{-5}$/  \\
   & 2.2 & BB & 0.55$\pm$0.04 & - & - & & 2.54 & 19 &  & & & 0.11\\
   & 1.05$\pm$0.53 & Brems. & 9.89$\pm$6.20 & - & - & & 0.72 & 18 &  & & & \\
   & 1.50$\pm$0.54 & RS & 5.40$\pm$1.37 & - & 1.57$\pm$1.15 & 0.18$\pm$0.09 & 0.66 & 16 &  & & & \\
   & 1.44$\pm$0.53 & MEKAL & 5.77$\pm$1.31 & - & 1.73$\pm$1.48 & 0.16$\pm$0.08 & 0.70 & 15  & & & \\
\hline

\hline

\end{tabular}

PL = power law, BB = blackbody, Brems. = bremsstrahlung, RS = Raymond Smith, Gau. = Gaussian line \\
$^*$ line centre = 1.00$\pm$0.05, $\sigma$ = 0.09$\pm$0.06 \hspace*{2.0cm}
$^@$ Errors are $\pm$0.02 and $\pm0.04$ respectively
\end{center}
\end{minipage}
\end{table*}

\subsection{Variability analysis}

We carried out a variability test on all the sources by dividing the
data into two equal frames of 17 ks and then into four equal frames of
8.5 ks and looking for variability in the source counts from frame to
frame.  However, within the errors, we found no significant
variability of any of the sources, unlike \cite{gend03a}, who found 
several variable sources in $\omega$ Centauri, using this method.

We also carried out a variability analysis on the nine sources with
enough counts for such a test.  These are the sources given in
Table~\ref{tab:xspectra}.  We extracted the lightcurves using regions
of the same size as those in the spectral analysis
(Sect.~\ref{sec:xspect}).  We used the filtered data between 0.2-10.0
keV that was uninterrupted by gaps due to the flare screening.  This
is important as gaps in the data can bias variability tests.  We used
the {\it ftool,lcstats}\footnote{http://heasarc.gsfc.nasa.gov/ftools/} \citep{blac95}
to perform a Kolmogorov-Smirnov probability of constancy test and the
$\chi^2$ probability of constancy test.  The results of these tests,
along with the size and number of data bins are given in
Table~\ref{tab:xspectra}.  Only one of the sources is significantly
variable, one of the core sources, source 39.

\section{The optical counterparts}

\begin{figure}
\hspace*{-1.2cm}     \includegraphics[angle=0,width=10.5cm]{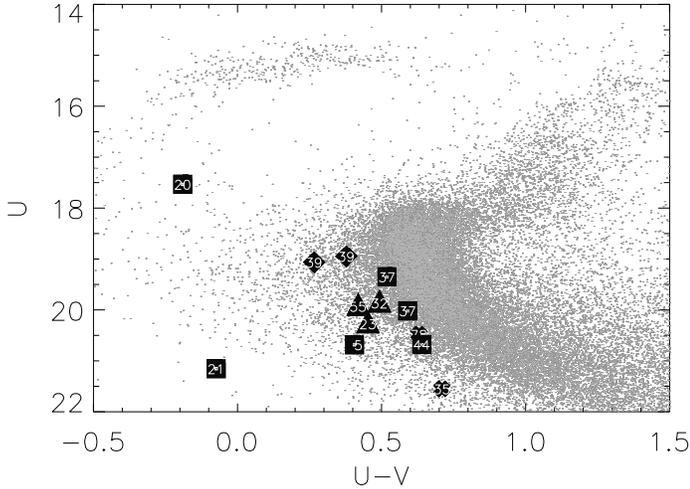}

     \caption{The U, U-V diagram of the globular cluster
     \object{M~22}.  The diamonds indicate blue optical sources found
     inside the error circle of an X-ray source in the core, the
     triangles indicate a blue optical source found inside the error
     circle of an X-ray source in the half mass radius, the squares
     indicate a blue optical source found inside the error circle of
     an X-ray source just outside the half mass-radius and the crosses
     indicate the blue optical sources found inside the error circle
     of an X-ray source way outside the half mass-radius (NB: source
     78 can just be seen behind source 44).  The X-ray source
     identification number (as given in Table~\ref{tab:xsources}) to
     which the blue source may be an optical counterpart is indicated
     on the symbol.  }  \label{fig:colormag}
\end{figure}

We constructed a U, (U-V) colour-magnitude diagram with the sources
from the central 12.5\arcmin\ of \object{M~22} (34\hspace*{0.5mm}614
sources), see Fig.\ref{fig:colormag}.  Only the central region was
used as it was only in this region that we found potential optical
counterparts of the X-ray sources (with the exception of source 35).
The majority of the optical sources in Fig.\ref{fig:colormag} are
likely to be cluster members, due to their centrally located
positions, although some of these sources may be foreground stars.
However, as all the data were taken at the same time, we have no
proper motion estimates for these sources and thus we can not discern
the cluster sources from foreground stars.  The average (U-V) value of
main-sequence stars in \object{M~22} is described by an arc given by
(U-V) = ((U/10.0)-1.3) $\pm$0.1 (22$<$U$<$18).  We have therefore
defined that a star is bluer than the main-sequence if (U-V) $<$
(U/10.0)-1.4.  We have identified blue sources that fall in the error
circle of the X-ray sources and they can be seen in the
colour-magnitude diagram in Fig.\ref{fig:colormag}, where the diamonds
indicate a blue optical source found inside the error circle of an
X-ray source in the core, the triangles indicate a blue optical source
found inside the error circle of an X-ray source in the half mass
radius, the squares indicate a blue optical source found inside the
error circle of an X-ray source just outside the half mass-radius and
the crosses indicate blue optical sources found inside the error
circle of an X-ray source way outside the half mass-radius.

We find that eleven X-ray source error circles contain at least one
blue optical source.  We find potential optical counterparts for one
of the core X-ray sources (39), all three X-ray sources in the
half-mass radius (23, 32 and 55) and three of the four X-ray sources
that fall just outside the half mass-radius (20, 21 and 34).  We also
find potential optical counterparts for three sources a few arcminutes
from the centre of the cluster (5, 44 and 78) and a possible optical
counterpart for source 35, which lies almost 14\arcmin\ from the
centre of the cluster.  We do not find an optical counterpart for the
two most central sources within the core, but this may be due to the
stellar confusion in this densely populated region.  In the cases of
sources 37 and 39, we find two possible optical counterparts.  The
positions as well as U and (U-V) magnitudes can be found in
Table~\ref{tab:opt_counterparts}.  Some of the astrometric matches
between the XMM-Newton X-ray sources and the blue stars maybe chance
coincidences.  We have calculated the number of blue stars per unit
area as a function of the radial distance from the centre of the
cluster, in the same way as \cite{edmo03}. We have thus been able to
calculate the probability of a chance coincidence.  We find
probabilities of 4\%, 0.6\% and 0.01\% in the core, half-mass radius
and outside the half mass radius.

As some of our X-ray sources could be active binaries, which have
red (cool) optical counterparts, not easily detectable with our optical
photometry, we have also cross-correlated several catalogues that
contain known variable stars in \object{M~22}
\citep{piet03,kalu01,clem01}.  However, none of the variable stars in
these catalogues fall within, or even close to, the error circles of
our X-ray sources.  The possible dwarf nova, detected by
\cite{ande03} {\it does} fall within the error circle of our X-ray source
36, which is {\it Rosat} source 4.  We discuss the nature of the X-ray
source in Sect.~\ref{sec:source23}.


\begin{table}
  \caption[]{Possible optical counterparts to the M22 X-ray sources}
     \label{tab:opt_counterparts}
       \begin{tabular}{ccccc}
         \hline
         \noalign{\smallskip}
X-ray & R.A. & Dec. & U & (U-V) \\
   Src. ID & $^h$ \hspace*{3mm} $^m$ \hspace*{3mm} $^s$ & $^{\circ}$ \hspace*{3mm} ' \hspace*{3mm} '' &  & \\         \hline
 20 & 18\ 36\ 15.6 & -23\ 51\ 02.4 & 17.53$\pm$0.02 & -0.19$\pm$0.03 \\
 32 & 18\ 36\ 19.0 & -23\ 52\ 18.0 & 19.86$\pm$0.03 & 0.49$\pm$0.05  \\    
 55 & 18\ 36\ 14.6 & -23\ 55\ 35.2 & 19.92$\pm$0.04 & 0.42$\pm$0.07  \\   
 35 & 18\ 35\ 30.2 & -23\ 54\ 33.0 & 21.55$\pm$0.06 & 0.70$\pm$0.07  \\    
 23 & 18\ 36\ 28.9 & -23\ 56\ 25.9 & 20.27$\pm$0.05 & 0.45$\pm$0.07  \\   
 39 & 18\ 36\ 25.4 & -23\ 55\ 20.8 & 18.94$\pm$0.05 & 0.38$\pm$0.08 \\   
 39 & 18\ 36\ 25.5 & -23\ 55\ 13.7 & 19.06$\pm$0.03 & 0.27$\pm$0.08  \\   
 37 & 18\ 36\ 38.7 & -23\ 54\ 42.5 & 19.34$\pm$0.03 & 0.52$\pm$0.05  \\   
 37 & 18\ 36\ 38.3 & -23\ 54\ 49.3 & 20.01$\pm$0.03 & 0.59$\pm$0.05 \\  
 21 & 18\ 36\ 45.4 & -23\ 51\ 58.4 & 21.15$\pm$0.05 & -0.07$\pm$0.08  \\   
  5 & 18\ 36\ 08.1 & -23\ 59\ 20.3 & 20.68$\pm$0.05 &  0.41$\pm$0.06  \\
 78 & 18\ 35\ 44.5 & -24\ 00\ 10.7 & 20.48$\pm$0.04 &  0.63$\pm$0.07  \\
 44 & 18\ 36\ 18.5 & -24\ 00\ 54.9 & 20.68$\pm$0.04 &  0.64$\pm$0.06  \\
\hline
  \end{tabular}
 \end{table}

\section{Discussion}

\subsection{Possible cluster members with fitted spectra}
\label{sec:poss_members}

\subsubsection{Source 36}
\label{sec:source23}

\begin{figure}
     \includegraphics[angle=-90,width=8cm]{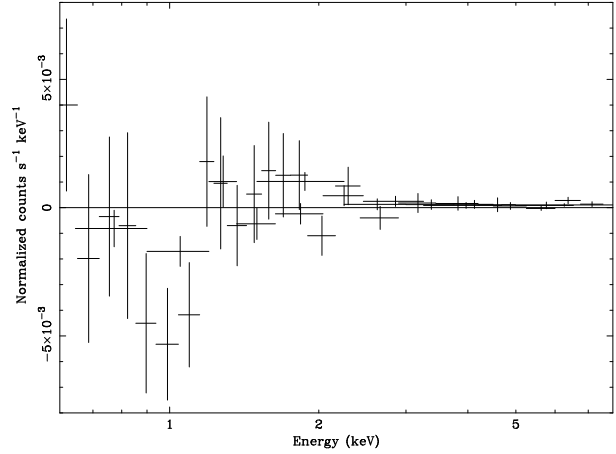}
     \caption{Residuals after fitting the EPIC spectrum of source 36
     with a simple power law model.  The region around 1 keV is
     clearly poorly fit by such a model.}  \label{fig:src54powresid}
\end{figure}                

\begin{figure}
     \includegraphics[angle=-90,width=8cm]{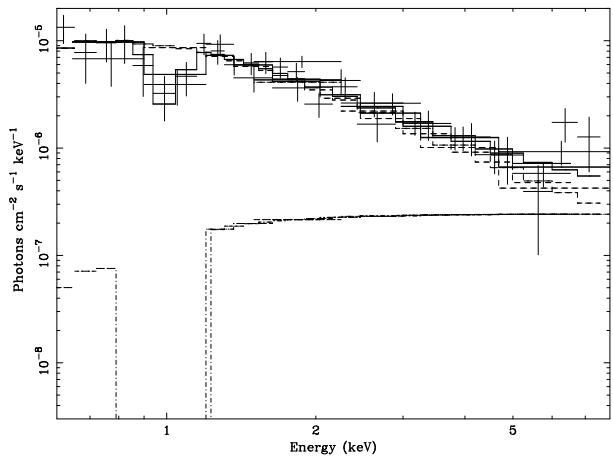}
     \caption{The EPIC spectrum for source 36, fitted with a power law
     fit and a Gaussian line (solid line).  The two components of this
     fit are shown as dashed lines. }  \label{fig:src54powgauss}
\end{figure}

This is the most centrally located source in the globular cluster
\object{M~22}.  Its absorption, see Table~\ref{tab:xspectra}, is 
consistent with that of the globular cluster.  The X-ray spectrum is
well fitted by a power law see Table~\ref{tab:xspectra}, although
other simple models such as a thermal bremsstrahlung give a good fit
to the data.  However, the data consistently over fit the region
around 1.0 keV, as can be seen in Fig.~\ref{fig:src54powresid}
where we have plotted the residuals resulting from the data minus the
power law fit.  Using a $\chi^{\scriptscriptstyle 2}$ test, we find
that the region about 1.0 keV deviates from the power law fit at
approximately the 2$\sigma$ level.  Thus we have tried to fit the
spectrum with a simple model (power law) and a Gaussian absorption
line.  This gives a good fit to the data, see Table~\ref{tab:xspectra}
and Fig.~\ref{fig:src54powgauss}.  We are confident that this is not
an instrumental effect, as there are no known instrumental edges in
the the region of this feature.  The absorption feature also appears
to be present in the data from all three cameras.  Such Gaussian lines
have been seen in the spectra of some neutron stars e.g. the neutron
star pulsar \object{1E1207.4-5209} \citep{sanw02,mere02,bign03} or
\object{RBS1223} (1RXS J130848.6+212708) \citep{habe03}, and are
thought to be due to resonant cyclotron absorption. \cite{bign03}
found three absorption lines at 0.7, 1.4 and 2.1 keV, embedded in the
continuum, which they fitted with two blackbodies (KT=0.21 and 0.40
keV).  Our data can also be well fitted with an absorption line as in
Table~\ref{tab:xspectra} and two blackbodies (KT$_1$=0.23$\pm$0.10 keV
and KT$_2$=0.74$\pm$0.13 keV, $\chi^{\scriptscriptstyle
2}_{\scriptscriptstyle \nu}$=1.17, 34 degrees of freedom).  This could
indicate that source 36 is similar in nature to \object{1E1207.4-5209}
or \object{RBS1223}.  If the absorption feature is real and due to
cyclotron resonance, we would expect a magnetic field strength of
approximately 1.1 $\times$ 10$^{11}$ G, if the feature is due to
electron cyclotron resonance, as the absorption lines observed in
\object{1E1207.4-5209}.  Alternatively  we would expect a magnetic 
field strength of approximately 2.1 $\times$ 10$^{14}$ G, if the
feature is due to proton cyclotron resonance, as the absorption line
observed in \object{RBS1223}.  We have also searched for a pulse
period between 0.001-10 secs, using Fourier analysis and period
folding.  However, we find no significant period in this range.  This
is not surprising, as such an object could be part of a binary system,
given the high stellar densities in globular clusters, especially in
the cores of GCs.  To detect periodicities of a millisecond pulsar in
a binary, it is imperative that the orbital motion is taken into
account.  Thus a pulsation could exist, but we are unable to detect
it, as we could be lacking information about a binary orbit.  A longer
observation may help to confirm these results, in particular to
identify whether there is a pulsation.  If this source is indeed a
neutron star, it is likely that it is a cluster member.

However, a Raymond Smith or a bremsstrahlung model can not be ruled
out as the most appropriate model to fit this data.  It is unlikely
that the Raymond Smith model is the best fit to the data if this
source is located in the globular cluster, see
Sect.~\ref{sec:xsources}, as this model is indicative of emission
from hot, diffuse gas, such as that from a galaxy.  However, a
bremsstrahlung model, with a temperature of $\sim$9 keV, gives a good
description of the data, discarding the possible absorption feature.
Such a model is indicative of a cataclysmic variable (dwarf nova) see
e.g. \cite{rich96,gend03a}, where we expect to find several such
sources in the cores of globular clusters.
\cite{ande03} found an optically variable source at
18$^h$36$^m$24$^s$66 and -23$^\circ$54\arcmin\ 35\arcsec5, which they
name CV1.  We do not detect this source in our optical data, probably
due to the stellar confusion in the centre of the core of the cluster.
They postulate that this source is a dwarf nova based on the star's
proper motion, H$_\alpha$ emission, variability and brightness in
quiescence (V$_{606}$ = 18.77) and they assert that the observed 1999
optical brightening ($\sim$ 3 magnitudes, Sahu et al. 2001) over 20-26
days was a dwarf nova outburst.  They also note that the source falls
within the X-ray error circle of the {\it Rosat} source 4 and thus
assert that this is the optical counterpart.  CV1 is located at the
edge of the positional error circle of our X-ray source 36.  However,
the light-curve of the alleged dwarf-nova outburst is rather
untypical. First, its long rise-time $\sim 10$ days is observed only
exceptionally, for example in so-called "anomalous" outbursts of the
dwarf nova SS Cyg \citep[][and references therein]{warn95}. Second,
its duration is much longer than that of normal outbursts and would
suggest a so-called superoutburst, but with a rather low amplitude
(observed nevertheless in the superoutburst prototype system
SU~UMa). In fact, the 1999 outburst light-curve is reminiscent of the
optical lightcurves of some X-ray transient event, especially from
low-mass binary systems containing neutron stars
\citep[see e.g.][]{chen97}. If the 1999 event were indeed related to
an X-ray outburst it would have been detected by RXTE ASM if its flux
was $\ga 2.3 \times 10^{-10}$ erg cm$^{-2}$ s$^{-1}$ or its luminosity
was $\ga 2.5 \times 10^{35}$ erg s$^{-1}$ (2-10 keV) at the distance
of the cluster. We have obtained the RXTE ASM lightcurve of this
source for the whole of 1999.  We find no evidence for a burst at the
time of the proposed outburst of CV1.  However, many cataclysmic
variable (CV) X-ray outbursts are fainter than this limit e.g. SS~Cyg
\citep{jone92} as are some X-ray binary outbursts \citep[see
e.g.][]{chen97}.  Therefore these observations can neither confirm nor
negate the outburst hypothesis.  The association of the 1999 outburst
with a neutron-star binary, where in this case the neutron star would
be a pulsar, could also imply that the best spectral fit is the power
law as in the case of other transient X-ray binaries containing
pulsars e.g. IGR J16358-4726 \citep{pate04} or SAX J1808.4-3658
\citep{camp02}, with a possible Gaussian absorption feature that could
be attributed to cyclotron resonant absorption. Further
multiwavelength observations should confirm the true nature of this
source.

\subsubsection{Source 39}
\label{sec:source39}

\begin{figure}
     \includegraphics[angle=-90,width=8cm]{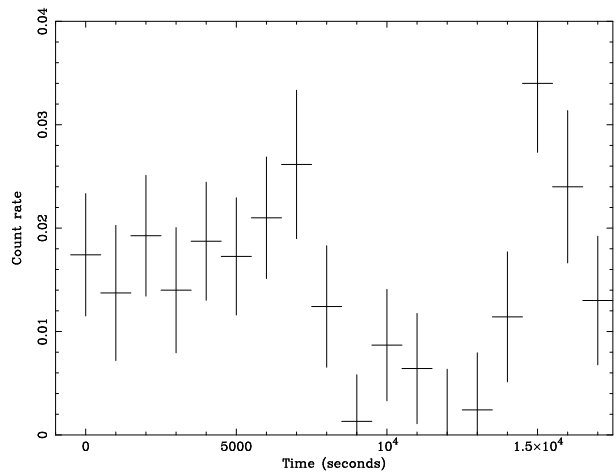} \caption{The PN
     data lightcurve (binsize=1000s) for source 39.  }
     \label{fig:src31lc}
\end{figure}

This core source is well fitted by a hard power law or by a high
temperature thermal bremsstrahlung, see Table~\ref{tab:xspectra} and
is found to be variable on timescales of hours, see
Fig.~\ref{fig:src31lc} and Table~\ref{tab:xspectra}.  We find two
blue optical sources in the error circle of the X-ray source with
U-band magnitudes of 18.9 and 19.1 and U-V of 0.4 and 0.3
respectively, see Table~\ref{tab:opt_counterparts}.  If either of
these is the optical counterpart, we find an L$_x$/L$_{opt}$ value of
$\sim$0.6, which is similar to that found for cataclysmic variables
\cite[e.g.][]{schw02,verb97}.  Both the spectrum and the temporal
analysis is consistent with this hypothesis, where cataclysmic variables
have high temperature bremsstrahlung spectra
\cite[e.g.][]{pool02,gend03a}.  Follow up spectroscopy should confirm 
this hypothesis and identify the optical counterpart.

\subsubsection{Source 33}
\label{sec:source33}

The faintest core source is best fitted by a two temperature (1
$\times$ 10$^6$ K and 4 $\times$ 10$^7$ K) raymond smith model, see
Table~\ref{tab:xspectra}.  \cite{demp93} found, following the analysis
of 44 active (RS~CVn) binaries observed with {\it Rosat}, that a two
temperature thermal plasma model, with typical temperatures of 2
$\times$ 10$^6$ K (with a scatter of approximately 1-3 $\times$ 10$^6$
K) and 1.6 $\times$ 10$^7$ K (with a scatter of approximately 1-4
$\times$ 10$^7$ K) gave a good description of the RS~CVn X-ray
spectra.  The fact that we find no optical counterpart to this source
in our photometric dataset could support the hypothesis that this
source is an RS~CVn binary, where our photometric dataset is designed
to detect blue sources and the optical counterparts of RS~CVn and
BY~Dra binaries are red \citep{demp97}.  Source 33's luminosity (4
$\times$ 10$^{31}$ erg s$^{-1}$ (0.1-2.4 keV) at the distance of the
cluster) is also consistent with the RS~CVn binary hypothesis
\citep{demp93}.  

One would perhaps expect that the X-ray light curve of an active
binary would be variable \citep[see e.g.][]{gend03a}, although it is
quite probable that during the full 9 hours of this observation no
discernible flickering or flaring event occurred, e.g.
\cite{oste00}.

\subsection{Possible cluster members without fitted spectra}
\label{sec:poss_membersnospec}

\subsubsection{Sources 5, 37 and 55}

Source 37 appears to be well fitted by a low temperature blackbody,
see Fig.~\ref{fig:xhard}.  Such a soft spectrum can indicate a
quiescent low mass X-ray binary with a neutron star primary, or a
foreground star \citep[see e.g.][]{gend03a,gend03b}.  We can
rule out the foreground star hypothesis as the radius of such a star
with the observed blackbody emission would be several orders of
magnitude smaller than the smallest M-dwarf stars.  From its position,
just outside the half mass-radius, this source could be a member of
the globular cluster and thus a quiescent low mass X-ray binary with a
neutron star primary.  However, its luminosity would then be $1 \times
10^{31} $erg s$^{-1}$, at the lower end of the faintest known
quiescent low mass X-ray binary with a neutron star primary,
\object{SAX J1808.4-3658} \citep[$5 \times 10^{31} $erg
s$^{-1}$][]{camp02}.  Also, if such systems are formed primarily
through collisions, as oppose to from their primordial binaries and
thus the relationship between the collision rate and number of neutron
star X-ray binaries \citep{gend03b,pool03,hein03} is valid, we do not
expect to find any such objects in \object{M 22}, due to its low
central density and thus low collision rate.

Such a soft spectrum can also be indicative of a millisecond
pulsar e.g. \cite{grin02}, where a blackbody of KT=0.5 keV indicates a
temperature of 5.8$\times 10^{6} $K, which is
consistent with the temperature of  the heated polar caps of a
millisecond pulsar \citep[10$^6$-10$^7$ K e.g.][and references
therein]{zhan03,zavl98}. Calculating the radius of the emission area
using the blackbody model fit, we find a radius of
0.05 km, which
is  smaller than the expected radius of emission from polar
caps \citep[$\sim$1 km e.g.][and references therein]{zhan03,zavl98}.
\cite{zavl96} state however, that 
blackbody models can produce higher temperatures and smaller sizes due
to the fact that the X-ray spectra emerging from light-element
atmospheres are harder than blackbody spectra. Indeed \cite{grin02}
found that the radii determined from the blackbody temperatures of the
millisecond pulsars in the globular cluster \object{47~Tuc} where also
significantly less than the expected radius of 1 km.  The X-ray
luminosity is also similar to field millisecond pulsars
\citep[e.g.][]{webb03}.  If source 37 is indeed a millisecond pulsar,
the optical counterpart suggested in Fig.~\ref{fig:colormag} may not
be the optical counterpart.  Millisecond pulsars are very faint in the
optical and could not be detected in the optical observations
presented here.  Many millisecond pulsars have companions
e.g. \cite{lund96}, but they can also be cool red stars, rather than
the hot blue stars, which our optical dataset is designed to detect.
However, the companion may be close enough that some accretion onto
the compact object continues, in the same way as in the accreting
millisecond pulsar \object{SAX J1808.4-3658}
\citep{mars98}, which could account for the blue nature of the
counterpart.  Alternatively, the companion star maybe constantly
illuminated by the pulsar, as in \object{PSR 1957+20}
\citep{fruc88}, which could also account for the blue nature of the
counterpart. From the X-ray colours, sources 5 and 55 may also be
similar objects.

\subsubsection{Sources 23, 44 and 32}

\begin{figure}
     \includegraphics[angle=0,width=9cm]{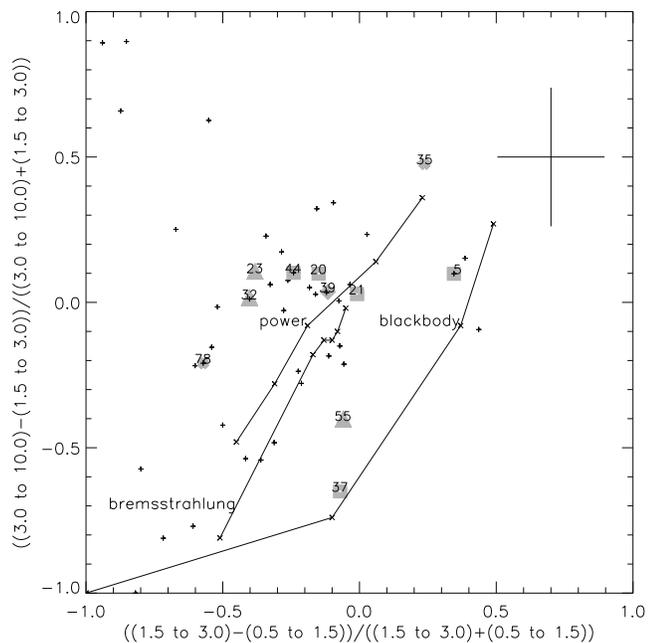} \caption{An X-ray
     colour diagram of all the detected X-ray sources in M~22.
     Pluses indicate an X-ray source.  Special attention is given to
     those sources for which a possible optical counterpart has been
     found.  The symbols are the same as those in
     Fig.~\ref{fig:colormag}.  The X-ray source identification
     number is that given in Table~\ref{tab:xsources}. Three lines are
     indicated showing the colours of a source in the cluster with a
     power law spectrum, a bremsstrahlung spectrum and a blackbody
     spectrum.  On the power law spectrum line, the crosses (from
     bottom to top) represent photon indices of 2.5, 2.0, 1.5, 1.0 and
     0.5.  On the bremsstrahlung spectrum line, the crosses (from
     bottom to top) represent temperatures of 1.0, 5.0, 10.0, 15.0,
     20.0 and 50.0 keV.  On the blackbody spectrum line, the crosses
     (from bottom to top) represent temperatures of 0.1, 0.5, 1.0, 1.5
     keV.  For N$_H$ values greater than that of the cluster, the
     spectral fits move towards the top right.  A typical error bar is
     also shown in the top right hand corner.}  \label{fig:xhard}
\end{figure}

The majority of the proposed optical counterparts (39, 37, 55, 32, 44,
5 and 23) have U-magnitudes between approximately 19 and 20.6 and
(U-V) values between 0.35 and 0.6.  These correspond to
6.6 $<$M$_U<$8.2 and 6.2 $<$M$_V<$7.6, values similar to those for
cataclysmic variables \cite[see e.g.][]{thor03,szko03}.  These sources
have unabsorbed X-ray fluxes of 8.6$\times$10$^{-15}$ -
1.2$\times$10$^{-13} {\rm ergs\ cm}^{-2} {\rm s}^{-1}$ (0.5-10.0 keV)
or 5.4$\times$10$^{-15}$ - 7.3$\times$10$^{-14} {\rm ergs\ cm}^{-2}
{\rm s}^{-1}$ (0.1-2.4 keV) using PIMMS (Mission Count Rate Simulator)
Version 3.4 and a 5 keV bremsstrahlung model.  Such values are
consistent with those of known cataclysmic variables fluxes.  Using a
colour-colour diagram to study the X-ray spectral nature of these
sources, as the majority do not have enough counts to be able to
extract and fit spectra, we find that their X-ray colours, with the
exception of sources 5, 37 and 55 which are discussed above, are also
similar to known cataclysmic variables.  Source 39, which is in the
centre of all these sources in Fig.~\ref{fig:xhard}, is well fitted by
a high temperature bremsstrahlung (see Sect.~\ref{sec:source39} and
Table~\ref{tab:xspectra}).  Sources 23, 44 and 32 appear to have
spectra that are more absorbed than the others.  This is nonetheless
consistent with a CV spectrum, as CV spectra can be absorbed due to
the high intrinsic absorption of the X-rays from the inner disc and/or
white dwarfs passing through the edge-on accretion disc e.g. one of
the CVs in \object{M 80} \citep{hein03}. All these sources are found
in the central regions of the cluster, where we expect to find such
binaries, due to mass segregation \citep[e.g.][]{meyl97}.  Thus if
these are the optical counterparts to the X-ray sources, they are
likely to be cataclysmic variables and thus members of the cluster.

\subsubsection{Sources 35 and 78}

Sources 35 and 78 have similar colours and luminosities to the
aforementioned sources, but they are found towards the edge of the
cluster, where we do not expect to find such binaries.  However,
several close binaries have been found in the outer regions of the
globular $\omega$ Centauri see e.g. \cite{gend03a}.  This has been
explained in several ways, including the idea that such binaries may
have been ejected through three-body interactions.  Thus it is
possible that these sources could also be cataclysmic variables and
members of the cluster.

\subsubsection{Source 21}

The two other possible optical counterparts in
Fig.~\ref{fig:colormag} have U, (U-V) magnitudes which are quite
different to the other possible optical counterparts.  Source 21 has a
faint enough U-magnitude that it could lie at the top of the white
dwarf cooling curve see e.g. \cite{knig02}, with a temperature of
approximately 80,000 K.  Its (U-V) value is also consistent with a
white dwarf hypothesis \cite[see e.g.][]{ches93}. However, the X-ray
spectrum is too hard for it to be a single white dwarf, see
Fig.~\ref{fig:xhard}.  Single white dwarfs can not have X-ray spectra
which peak later than 0.5 keV \citep[see][]{odwy03}, thus to have such
a hard X-ray spectrum, the white dwarf must have a companion.  This
companion could either be a late-type active star which is too faint
to be observed in the U, B or V, but the stellar activity can be
detected in X-rays.  Alternatively, the hard X-ray emission can be
produced through accretion onto the white dwarf from a companion again
too faint and/or red to be detected in the optical data, in which case
this too is a cataclysmic variable \citep[see e.g.][]{odwy03}.

\subsubsection{Source 20}

Source 20 is also interesting due to its position on the
colour-magnitude diagram, see Fig.~\ref{fig:colormag}.  It lies
below the blue horizontal branch, home to post-main sequence stars, in
the region where we expect blue straggler stars.  Blue stragglers are
thought to be formed through collisions and thus are more readily
formed in globular clusters, where the stellar densities are high.
Other blue stragglers have also been detected in X-rays
e.g. \cite{bell98,bell96}, however in all cases, the `hard', faint
X-ray emission has been due to a companion star.

\subsubsection{Source 34}

Source 34 is one of the sources that is clustered in the bottom
left-hand corner of Fig.~\ref{fig:xhard}.  These are the softest
sources in the cluster.  The other two sources are 31 and 17, where
source 31 is, from spectral fitting, its extent and location (towards
the edge of the field of view) likely to be a background source.
Source 17 is also likely to be a background source from its extent and
location.  However, source 34 is situated very close to the half mass
radius and thus could be related to the cluster, in which case, with a
luminosity of $3 \times 10^{31} $erg s$^{-1}$ it could be a quiescent
low mass X-ray binary with a neutron star primary or a millisecond
pulsar, as discussed above for source 37.  Alternatively, this
source could be an active binary in the cluster, with temperatures of
approximately $1 \times 10^{6}$ and $1 \times 10^{7}$K.  Its
luminosity (3$\times$10$^{31}$ erg s$^{-1}$ (0.1-2.4 keV) at the
distance of the cluster) and lack of optical counterpart could support
this theory.  From its temperature, this source could also be a late-type
active foreground star, but we can rule out this hypothesis as its
radius would then be several orders of magnitude smaller than the smallest
M-dwarf stars.

\subsection{Non cluster members with fitted spectra}
\label{sec:nonposs_members}

\subsubsection{Source 30}

\begin{figure}
     \includegraphics[angle=-90,width=8cm]{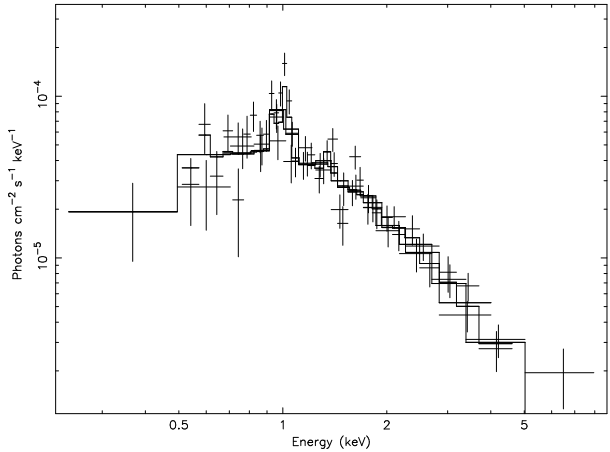}
     \caption{The EPIC spectrum for source 30, fitted with a MEKAL model.}  \label{fig:src45mekalspec}
\end{figure}

This extended source was detected previously by {\it Einstein}
\citep{hert83} and {\it Rosat} but with too few counts to be
identified.  From its distance from the cluster centre and its extent,
source 30 is likely to be a background source.  Indeed, its X-ray
spectrum is best fitted by a MEKAL model, see Table~\ref{tab:xspectra}
and Fig.~\ref{fig:src45mekalspec} or a Raymond Smith model, indicative
of emission from a hot diffuse system.  Using the distance obtained
from the spectral fitting and the extent of the object in our data, we
can calculate the radius of the object, which is 0.2-0.4Mpc.  This
indicates that it could be a giant elliptical galaxy or a cluster of
galaxies.  However, the temperature is too high for it to be a giant
elliptical \citep{osul01}.  The luminosity (0.9-3.7$\times$10$^{43}$
erg s$^{-1}$) and temperature are in good agreement, however, with the
luminosity-temperature relationship known for clusters of galaxies,
e.g. \cite{babu02}, thus it is likely that this source is a background
object and a galaxy cluster.  This is in agreement with \cite{hert83},
who proposed that due to its extent and distance from the cluster
core, it could be either a supernova remnant or a cluster of
galaxies.  They ruled-out the supernova remnant hypothesis using the fact
that the probability of observing such a remnant by chance projection
in any field of their survey was $<$0.05, whereas to detect a cluster
of galaxies, the probability was more than ten times more likely.
They determine a similar luminosity to that determined here,
($\sim$0.5-1.0$\times$10$^{44}$ erg s$^{-1}$), using the assumption
that the flux that they observed emanates from the central 500 kpc
core of the cluster.

\subsubsection{Other possible background sources without fitted spectra}
\label{sec:bkg_src}

From the spectral fitting, the extent of the sources and their
location (towards the edge of the field of view), sources 31 and 29
are also likely to be background sources.  Sources 31 and 17 are both
located in the softest region of the X-ray colour-colour diagram,
Fig.~\ref{fig:xhard}, and as discussed above source 31 is, from
spectral fitting, its extent and location (towards the edge of the
field of view) likely to be a background source.  Source 17 is also
likely to be a background source from its extent and location.
Sources 42 and 67, which are found in the top left corner of the X-ray
colour-colour diagram, see Fig.~\ref{fig:xhard}, are also likely to
be a background sources, as their spectra a very hard, with few counts
at lower energies, indicating strong interstellar absorption and hence
an origin behind the globular cluster.  They are also found towards
the edge of the field of view, and thus it is unlikely that they
belong to the cluster. Other sources such as 75 and 60, which are
located in a similar place to sources 42 and 67 in the X-ray
colour-colour diagram and are also positioned towards the edge of the
field of view, are also likely to be background sources.

\section{Conclusions}

We have presented both X-ray (0.2-10.0 keV) and optical (U, B and V
photometry) of the globular cluster M~22.  We have shown that 5$\pm$3
sources belong to the cluster, where these sources are clustered about
the centre of the GC.  Using the broad band X-ray spectra and X-ray
timing analysis, we suggest that two of the three core sources, which
are likely to belong to the cluster are either a pulsar and a
cataclysmic variable or two cataclysmic variables.  We find probable
optical counterparts for as many as 11 sources.  Using the X-ray and
optical data we have presented evidence to show that other sources
maybe cataclysmic variables, millisecond pulsars, active
binaries or blue stragglers.

The majority of the sources detected in the field of view are
background objects.  We provide evidence for a cluster of galaxies
situated behind M~22 as well as several other galaxies.  Planned
follow-up spectroscopy of these sources with the VLT UT3 (VIMOS) will
help clarify the uncertain nature of these sources.

\begin{acknowledgements}
We are indebted to the authors of the WFPRED package at Padova
University, in particularly to E. Held. We acknowledge M. Watson for
his advice on source 30.  We also thank the anonymous referee for
their useful comments and suggestions.  This article was based on
observations obtained with XMM-Newton, an ESA science mission with
instruments and contributions directly funded by ESA Member States and
NASA.  We are also grateful to R. Remillard and all the RXTE ASM team
for providing us with the data concerning source 36.  The authors NAW
and DB also acknowledge the CNES for its support in this research.
\end{acknowledgements}

\end{document}